\documentstyle[12pt]{article}

\author{V.A.Tsokur and Yu.M.Zinoviev
       \thanks{E-mail address: ZINOVIEV@MX.IHEP.SU} \\
        {\it Institute for High Energy Physics} \\
        {\it Protvino, Moscow Region, 142284, Russia}}
\title{$N=2$ Supergravity Models\\
        Based on the Nonsymmetric Quaternionic Manifolds \\
       II. Gauge Interactions}
\date{March 1996}

\begin{document}

\maketitle

\begin{abstract}
   In this paper we continue our investigation of the $N=2$
supergravity models, where scalar fields of hypermultiplets
parameterize the nonsymmetric quaternionic manifolds. Using the
results of our previous paper, where we have given an explicit
construction for the Lagrangians and the supertransformations and,
in-particular, the known global symmetries of the Lagrangians, we
consider here the switching on the gauge interaction. We show that in
this type of models there appears to be possible to have spontaneous
supersymmetry breaking with two different scales and without a
cosmological term. Moreover, such a breaking could lead to the
generation of the Yukawa interactions of the scalar and spinor fields
from the hypermultiplets which are absent in other known models.
\end{abstract}

\newpage

\section*{Introduction}

   In our previous paper \cite{Tso96} we have considered the $N=2$
supergravity models where scalar fields of hypermultiplets
parameterize one of two types \cite{Ale75,Cec89} of the
non-symmetric quaternionic manifolds. We have managed to give an
explicit construction of the appropriate Lagrangians and
supertransformations in terms of the usual hypermultiplets. One of
the important general features of these models is the fact that all
of them contain as a common component one of the three possible
hidden sectors \cite{Zin90}, admitting spontaneous supersymmetry
breaking with two arbitrary scales and without a cosmological term.
So, in this paper we consider the possibility to switch on the gauge
interactions in such models, paying the main attention to the ones
which lead to the spontaneous supersymmetry breaking. For the vector
multiplets we choose the well known model with the scalar field
geometry $O(2,m)/O(2)\otimes O(m)$. The first reason is that this
model appears to be the natural generalization of the $N=2$ hidden
sector \cite{Cec86a,Zin86} to the case of the arbitrary number of
vector multiplets as it was shown in \cite{Zin90}, where the usual
symmetric quaternionic manifolds were investigated. The second one is
that such a model arises in the low-energy limit of four-dimensional
superstrings with $N=2$ supersymmetry. For completeness we reproduce
all the necessary formulas below. As it was expected, both types of
models do admit the spontaneous supersymmetry breaking with two
different scales and without a cosmological term, in this a number of
soft breaking terms arises as a result of this breaking. As we will
show, apart from the mass terms for different scalar and spinor
fields of vector and hypermultiplets, in one type of models the
supersymmetry breaking leads to the appearance of the Yukawa
interactions between scalar and spinor fields of hypermultiplets.

\section{Vector multiplets}

To describe the interaction of vector multiplets with N=2
supergravity, let us introduce the following fields: graviton $e_{\mu
r}$, gravitini $\Psi_{\mu i}$, $i=1,2$, Majorana spinors $\rho_i$,
scalar fields $\hat{\varphi}$, $\hat{\pi}$, and $(m+2)$ vector
multiplets $\{A_\mu^M, \Theta_i^M,{\cal Z}^M ={\cal
X}^M+\gamma_5{\cal Y}^M \}$, $M=1,2,...m+2$, $g^{MN}=(--,+...+)$. It
is not difficult to see that the set of spinor and scalar fields is
superfluous (which is necessary for symmetrical description of
graviphotons and matter vector fields). The following set of
constraints corresponds to the model with the geometry
$O(2,m)/O(2)\otimes O(m)$: 
\begin{equation}
\bar{\cal Z}\cdot{\cal Z}=-2, \qquad {\cal Z}\cdot{\cal Z}=0,  \qquad 
{\cal Z}\cdot\Theta_i=\bar{\cal Z}\cdot\Theta_i=0.    \label{10}
\end{equation}

    The number of the physical degrees of freedom is correct only
when the theory is invariant under the local $O(2) \approx  U(1)$
transformations, the combination $(\bar{\cal{Z}}\partial_\mu{\cal
Z})$ playing the role of a gauge field. Covariant derivatives for
scalar fields ${\cal Z}$ and $\bar{\cal{Z}}$ look like
\begin{equation} 
 D_{\mu} = \partial_{\mu} \pm \frac12 (\bar{\cal{Z}} \partial_{\mu}
{\cal Z}), \label{11}
\end{equation}
where covariant derivative $D_{\mu}{\cal Z}$ has the sign "+" and
$D_{\mu}\bar{\cal Z}$ has the sign "-".

    In the given notations the Lagrangian of interaction looks as
follows:
\begin{eqnarray}
 {\cal L}^F &=& \frac{i}2 \varepsilon^{\mu\nu\rho\sigma}
\bar{\Psi}_{\mu i} \gamma_5 \gamma_\nu D_\rho \Psi_{\sigma i} +
\frac{i}2 \bar{\rho }^i \hat{D} \rho_i + \frac{i}2 \bar{\Theta }^i
\hat{D} \Theta_i - \nonumber \\
 && + e^{\hat\varphi /\sqrt{2}} \left\{ \frac14 \varepsilon^{ij}
\bar{\Psi}_{\mu i} ({\cal{Z}} (A^{\mu\nu} - \gamma_5
\tilde{A}^{\mu\nu})) \Psi_{\nu j} + \frac14 \bar{\Theta }^i
\gamma^\mu (\sigma A) \Psi _{\mu i} + \right. \nonumber \\
 && + \left. \frac{i}{4\sqrt{2}} \bar{\rho}^i \gamma^\mu ({\cal Z}
(\sigma A)) \Psi_{\mu i} + \frac{\varepsilon^{ij}}8 \left[ 2\sqrt{2}
\bar{\rho }_i (\sigma A) \Theta_j + \bar{\Theta}_i{}^M ({\cal Z}
(\sigma A)) \Theta_j{}^M \right] \right\}     \nonumber   \\
 && - \frac12 \varepsilon^{ij} \bar{\Theta}_i{}^M \gamma^\mu
\gamma^\nu D_\nu {\cal Z}^M \Psi_{\mu j} - \frac12 \varepsilon^{ij}
\bar{\rho}_i \gamma^\mu \gamma^\nu (\partial_\nu  \hat{\varphi}  +
\gamma_5 e^{-\sqrt{2}\hat{\varphi}} \partial_\nu \hat{\pi}) \Psi_{\mu
j} \label{12}    \\
 {\cal L}^B &=& - \frac12 R - \frac14 e^{\sqrt{2}\hat{\varphi}}
\left[ A_{\mu\nu}{}^2 + 2 ({\cal Z} \cdot A_{\mu\nu}) (\bar{\cal Z}
\cdot A_{\mu\nu}) \right] - \frac{\hat{\pi}}{2\sqrt{2}} (A \cdot
\tilde{A}) +   \nonumber \\
 && + \frac12 (\partial_\mu \hat{\varphi})^2 + \frac12
e^{-2\sqrt{2}\hat{\varphi}} (\partial_\mu \hat{\pi})^2 + \frac12
D_\mu {\cal Z}^A D_\mu \bar{\cal Z}^A.  \label{15}
\end{eqnarray}

Derivatives of the spinor fields have the following form:
\begin{eqnarray}
 D_\mu \eta_i &=& D_\mu^G \eta_i - \frac14 (\bar{\cal Z} \partial_\mu
{\cal Z}) \eta_i + \frac1{2\sqrt{2}} e^{-\sqrt{2}\hat{\varphi}}
\gamma_5 \partial_\mu \hat{\pi} \eta_i,     \nonumber   \\
 D_\mu \rho_i &=& D_\mu^G \rho_i + \frac14 (\bar{\cal Z} \partial_\mu
{\cal Z}) \rho_i + \frac3{2\sqrt{2}} e^{-\sqrt{2}\hat{\varphi}}
\gamma_5 \partial_\mu \hat{\pi} \rho_i,       \label{13}    \\
 D_\mu \Theta_i &=& D_\mu^G \Theta_i - \frac14 (\bar{\cal
Z}\partial_\mu {\cal Z}) \Theta_i - \frac1{2\sqrt{2}}
e^{-\sqrt{2}\hat{\varphi}} \gamma_5 \partial_\mu \hat{\pi} \Theta_i,
\nonumber
\end{eqnarray}
and the derivative of the field $\Psi_{\mu i}$ is the same as for
$\eta_i$.

Supertransformation laws look like
\begin{eqnarray}
 \delta \Theta^M_i &=& - \frac12 e^{\hat{\varphi}/\sqrt{2}}
\left\{ (\sigma A)^M + \frac12 \bar{\cal Z}^M ({\cal Z} (\sigma A)) +
\frac12 {\cal Z}^M (\bar{\cal Z} (\sigma A)) \right\} \eta_i - i 
\varepsilon_{ij} \hat{D} {\cal Z}^M \eta_i,        \nonumber   \\
 \delta \rho_i &=& - \frac1{2\sqrt{2}} e^{\hat{\varphi}/\sqrt{2}}
{\cal Z} (\sigma A) \eta_i - i \varepsilon_{ij} \gamma^\mu
(\partial_\mu \hat{\varphi} + \gamma_5 e^{-\sqrt{2}\hat{\varphi}}
\partial_\mu \hat{\pi}) \eta_i,          \nonumber  \\
 \delta \Psi{\mu i} &=& 2 D_\mu \eta_i + \frac{i}4 \varepsilon_{ij}
e^{\hat{\varphi}/\sqrt{2}} \bar{\cal Z} (\sigma A) \eta_i    \qquad
\delta \hat{\pi} = e^{\sqrt{2}\hat{\varphi}} \varepsilon^{ij}
(\bar{\rho}_i \gamma_5 \eta_j),            \nonumber     \\
 \delta {\cal X}^A &=& \varepsilon^{ij} (\bar{\Theta}_i{}^A \eta_j)
\qquad \delta {\cal Y}^A = \varepsilon^{ij} (\bar{\Theta}_i{}^A
\gamma_5 \eta_j)   \qquad \delta \hat{\varphi} = \varepsilon^{ij}
(\bar{\rho}_i \eta_j),    \nonumber    \\
 \delta A_\mu^A &=& e^{-\hat{\varphi}/\sqrt{2}} \left\{
\varepsilon^{ij} (\bar{\Psi}_{\mu i} {\cal Z}^A \eta_j ) + i
(\bar{\Theta}_i^A \gamma_\mu \eta^i) - \frac{i}{\sqrt{2}}
(\bar{\rho}^i \gamma_\mu {\cal Z}^A \eta_i) \right\}.   \label{16} 
\end{eqnarray}

\section{$W(p,q)$-model}

The W(p,q)-model has been constructed in \cite{Tso96} and here we
shall give only a brief description of it. The main attention will be
paid to the switching on a gauge interaction in this model and to the
investigation of the spontaneous supersymmetry breaking and its
consequences.

\subsection{Description of the model}

  To describe W(p,q)-model let us introduce two kinds of the
hypermultiplets $(\Lambda_\alpha, Y_m)^{\dot{A}}$ and
$(\Sigma_\alpha, Z^m)^{\ddot{A}}$, $\dot{A} = 1,...,p$, $\ddot{A} =
1,...,q$, $\alpha=1,2$ and $m=1,2,3,4$, which we call correspondingly
Y- and Z-multiplets. These multiplets interact with a hidden sector
that has been constructed in \cite{Zin90} and contains the following
fields: graviton $e_{\mu r}$, gravitini $\Psi_{\mu i}$, $i=1,2$,
fermionic fields $\chi^a_\alpha$, $a=1,2,3,4$, and bosonic fields
$y_{ma}$ and $\pi^{[mn]}$. The fields $\pi^{mn}$ will enter the
Lagrangian through a derivative only, whereas $y_{ma}$ will realize
the nonlinear $\sigma$-model $GL(4,R)/O(4)$. 

Let us denote $y_{ma}^{-1}$ as $y^{ma}$, so that $y_{ma} y^{na} = 
\delta^n_m$.

We shall need four matrices $(\tau^a)_{i\alpha}$ and their conjugate
ones $(\bar{\tau}^a)^{\alpha i}$ satisfying the condition:
\begin{equation}
\tau^a\bar\tau^b+\tau^b\bar\tau^a=2\delta^{ab}I,    \label{60}
\end{equation}
for which we shall use the following explicit representation $\tau = 
\{I,\vec{\sigma}\}$, $\bar{\tau} = \{I,-\vec{\sigma}\}$, where
$\vec{\sigma}$ are the Pauli matrices. Let us introduce also six
matrices:
\begin{equation}
\Sigma^{ab}=\frac{1}{2}(\tau^a\bar\tau^b-\tau^b\bar\tau^a),  \qquad
\Sigma^{ab}=\frac{1}{2}\varepsilon^{abcd}\Sigma^{cd}.     \label{61}
\end{equation}
Recall, since in our representation for spinors the matrix $\gamma_5$
plays the role of an imaginary unit, then, e.g.,
\begin{equation}
\gamma_\mu(\tau^a)_{i\alpha}=(\bar\tau^a)^{\alpha i}\gamma_\mu,
\qquad \gamma_\mu(\Sigma^{ab})_i{}^j=-(\Sigma^{ab})^i{}_j\gamma_\mu.
\label{62}   
\end{equation}

The Lagrangian of the interaction of the hidden sector with Y- and
Z-multiplets have the following form:
\begin{eqnarray}
{\cal L^F} &=& \frac{i}2 \varepsilon^{\mu\nu\rho\sigma} 
\bar\Psi_{\mu i} \gamma_5 \gamma_{\nu} D_{\rho} \Psi_{\sigma i} +
\frac{i}2 \bar\chi^a_\alpha \hat D \chi^a_\alpha + \frac{i}2
\bar\Lambda_\alpha \hat D \Lambda_{\alpha} + \frac{i}2
\bar\Sigma_\alpha \hat D \Sigma_{\alpha} -  \nonumber   \\
 && - \frac12 \bar\chi^a_\alpha \gamma^\mu \gamma^\nu (S_\nu^+ +
P_\nu)_{ab} (\bar \tau^b)^{\alpha i} \Psi_{\mu i} - \frac12
\bar\Sigma_\alpha \gamma^\mu \gamma^\nu \partial_\nu Z^m y_{ma}
(\bar\tau^a)^{\alpha i} \Psi_{\mu i} -    \nonumber  \\
 && - \frac12 \bar\Lambda_\alpha \gamma^\mu \gamma^\nu
\sqrt\Delta\partial_\nu Y_m y^{ma} (\bar\tau^a)^{\alpha i} 
\Psi_{\mu i} - \frac{i}2 \bar\chi^a_\alpha \gamma^\mu (S_\mu^- -
P_\mu)_{ab} \chi^b_\alpha +     \nonumber    \\
 && + i \bar\Sigma_\alpha \gamma^\mu \partial_\mu Z^m y_{ma}
\chi^a_{\alpha} - \frac{i}2 \bar\Lambda_\alpha \gamma^\mu
\sqrt\Delta\partial_\mu Y_my^{ma} \chi^a_\alpha +    \nonumber    \\
 && + \frac{i}2 \bar\Lambda_\alpha \gamma^\mu \partial_\mu Y_m y^{mb}
(\Sigma^{ab})_\alpha{}^\beta \chi^a_\beta,     \label{63}       \\
 {\cal L^B} &=& - \frac12 R + \frac12 (S^+_{\mu ab})^2 + \frac12
(P_{\mu ab})^2 + \frac12 g_{mn} \partial_\mu Z^m \partial_\mu Z^n + 
 \frac{\Delta}2 g^{mn} \partial_\mu Y_m \partial_\mu Y_n,
\end{eqnarray}
where we have denoted:
\begin{eqnarray}
 P_{\mu ab} &=& y_{ma} (\partial_\mu \pi^{mn} + \frac12 Z^m \stackrel
{\leftrightarrow}{\partial_\mu} Z^n + \frac14 \varepsilon^{mnpq}
Y_p\stackrel{\leftrightarrow}{\partial_\mu} Y_q) y_{nb},  \nonumber
\\
 S_{\mu ab}^\pm &=& \frac12 (\partial_\mu y_{ma} y^{mb} \pm y_{mb}
\partial_\mu y^{ma})               \label{65}
\end{eqnarray}
and the D-derivatives of the fermionic fields look like:
\begin{equation}
D_\mu = D_\mu^G \pm \frac14 (S_\mu^-+P_\mu)_{ab} \Sigma^{ab}
\label{651}
\end{equation}
with the sign "-" for the derivatives of the parameter $\eta$ and
gravitino $\Psi_\mu$ and the sign "+" for all other fermion
fields derivatives.

Corresponding supertransformation laws have the following form:
\begin{eqnarray}
 \delta \Psi_{\mu i} &=& 2 D_\mu \eta_i   \qquad  \delta
\chi^a_\alpha = - i \gamma^\mu (S_\mu^+ + P_\mu)_{ab}
(\bar\tau^b)^{\alpha i} \eta_i,       \nonumber   \\
 \delta y_{ma} &=& y_{mb} (\bar\chi^b_\alpha (\bar\tau^a)^{\alpha i}
\eta_i),  \quad \delta \pi^{mn} = \frac12 \{y^{ma} y^{nb} - y^{mb}
y^{na} \} (\bar\chi^a_\alpha (\bar\tau^b)^{\alpha i}\eta_i),
\nonumber \\
 \delta \Lambda_\alpha &=& - i \gamma^\mu \sqrt\Delta \partial_\mu
Y_m y^{ma} (\bar\tau^a)^{\alpha i} \eta_i,   \quad \delta Y_m =
\frac1{\sqrt\Delta} y_{ma} (\bar\Lambda_\alpha (\bar\tau^a)^{\alpha
i} \eta_i),     \nonumber  \\
 \delta \Sigma_\alpha &=& - i \gamma^\mu \partial_\mu Z^m y_{ma}
(\bar\tau^a)^{\alpha i} \eta_i,  \quad \delta Z^m = y^{ma}
(\bar\Sigma_\alpha (\bar\tau^a)^{\alpha i} \eta_i)   \label{66}
\end{eqnarray}
with the notations defined above.

Now let us consider interaction of W(p,q)-model with the vector
multiplets described in the previous section. It is easy to check
that the only additional terms, which appear in the Lagrangian, are
the following:
\begin{eqnarray}
 \Delta {\cal L^F} &=& \frac{e^{\hat\varphi/\sqrt 2}}8
\varepsilon^{\alpha\beta} \left\{ (\bar\chi^a_\alpha \bar{\cal Z}^M
(\sigma A)^M \chi^b_\beta) + \right.  \nonumber     \\
 && + \left.( \bar\Lambda_\alpha \bar{\cal Z}^M (\sigma A)^M
\Lambda_\beta) + (\bar\Sigma_\alpha \bar{\cal Z}^M (\sigma A)^M
\Sigma_\beta)\right\}. \label{67}
\end{eqnarray}
Besides, D-derivatives of the fermionic fields change their form.
Derivatives (\ref{651}) of all the fermions of W(p,q)-model acquire
the following additional terms (analogously to the ones in
(\ref{13})):
\begin{equation}
\frac{1}{4}(\bar{\cal Z}\partial_\mu{\cal Z})-\frac{1}{2\sqrt 2}
e^{-\sqrt 2\hat\varphi}\gamma_5\partial_\mu\hat\pi   \label{68}
\end{equation}
and derivatives (\ref{13}) of the fields $\rho_i$ and $\Theta_i^M$
acquire the following additional terms (analogously to the ones in
(\ref{651})):
\begin{equation}
\frac{1}{4}(S_\mu^-+P_\mu)_{ab}(\Sigma^{ab})_i{}^j.
\label{69}
\end{equation}

\subsection{Gauge interaction and symmetry breaking}

Our next step will be to switch on the gauge interaction and
investigate a possibility to have a spontaneous supersymmetry
breaking and its consequences.

 Among the global symmetries of the bosonic Lagrangian there are the
translations of the field $\pi^{mn}$: $\pi\to\pi+\Lambda$. It has
been shown in \cite{Zin90}, that for the three out of these six
translations their gauging leads to the spontaneous supersymmetry
breaking with a vanishing cosmological constant. Also, a bosonic
Lagrangian of the model is invariant under the global transformations
of the group $O(p)\otimes O(q)$, which touches the Y- and Z-sectors,
and that allows one to switch on the gauge interaction corresponding
to some subgroup of this group. For that let us make the following 
substitution in the Lagrangian and the supertransformation laws:
\begin{eqnarray}
 \partial_\mu Y_m^{\dot A} &\to& \partial_\mu Y_m^{\dot A} - A_\mu^M
(T^M)^{\dot A\dot B} Y_m^{\dot B},   \quad \partial_\mu Z^{m\ddot A}
\to \partial_\mu Z^{m\ddot A} - A_\mu^M (T^M)^{\ddot A\ddot B}
Z^{m\ddot B},   \nonumber  \\
 \partial_\mu {\cal Z}^M &\to& \partial_\mu {\cal Z}^M - f^{MNK}
A_\mu{}^N {\cal Z}^K, \quad \partial_\mu \Theta_i{}^M \to
\partial_\mu \Theta_i{}^M - f^{MNK} A_\mu{}^N \Theta_i{}^K,
\nonumber    \\
 \partial_\mu \pi^{mn} &\to& \partial_\mu \pi^{mn} - A_\mu^M
(M^M)^{mn}.   \label{70}        
\end{eqnarray}

    In order to restore the invariance of the Lagrangian under the 
supertransformations, one has to add the following terms to the
Lagrangian 
\begin{eqnarray}
 {\cal L'^F} &=& e^{-\hat\varphi/\sqrt 2} \left\{ - \frac14
\bar\Psi_{\mu i} \sigma^{\mu\nu} \varepsilon^{ij} ({\cal Z} R)_{ab}
(\Sigma^{ab})_j{}^k \Psi_{\nu k} + \frac{i}{2\sqrt 2} \bar\Psi_{\mu
i} \gamma^\mu (\bar{\cal Z} R)_{ab} (\Sigma^{ab})_i{}^j \rho_j -
\right.     \nonumber    \\
 && - \frac{i}2 \bar\Psi_{\mu i} \gamma^\mu R_{ab}
(\Sigma^{ab})_i{}^j \Theta_j - i \bar\Psi_{\mu i} \gamma^\mu
\varepsilon_{ij} (\bar\tau^a)^{\alpha j} (\bar{\cal Z} R)_{ab}
\chi^b_\alpha -       \nonumber   \\
 && - \frac{i}2 \bar\Psi_{\mu i} \gamma^\mu \sqrt\Delta Y_m^{\dot A}
\bar{\cal Z}^{\dot A\dot B} y^{ma} (\tau^a)_{i\alpha}
\varepsilon^{\alpha\beta} \Lambda_\beta^{\dot B} - \frac{i}2
\bar\Psi_{\mu i} \gamma^\mu Z^{m\ddot A}\bar{\cal Z}^{\ddot A\ddot B}
y_{ma} (\tau^a)_{i\alpha} \varepsilon^{\alpha\beta}
\Sigma_\beta^{\ddot B}       \nonumber    \\
 && + \frac{1}{\sqrt2} \bar\rho_i \varepsilon^{ij} R_{ab}
(\Sigma^{ab})_j{}^k \Theta_k - \sqrt 2 \bar\rho_i (\bar{\cal Z}
R)_{ab} (\bar\tau^a)^{\alpha i} \chi^b_\alpha - \frac14
\bar{\Theta}_i{}^M \varepsilon^{ij} ({\cal Z} R) \Sigma_j{}^k
\Theta_k{}^M - \nonumber    \\
 && - \frac1{\sqrt 2} \bar\rho_i \sqrt\Delta Y_m^{\dot A} 
\bar{\cal Z}^{\dot A \dot B} y^{ma} (\bar\tau^a)^{\alpha i}
\Lambda_\alpha^{\dot B} - \frac1{\sqrt 2} \bar\rho_i Z^{m\ddot A}
\bar{\cal Z}^{\ddot A\ddot B} y_{ma} (\bar\tau^a)^{\alpha i}
\Sigma_\alpha^{\ddot B}  +     \nonumber   \\
 && + 2 \bar\Theta_i R_{ab} (\bar\tau^a)^{\alpha i} \chi^b_\alpha -
\bar\chi^a_\alpha \varepsilon^{\alpha\beta} (\bar{\cal Z} )_{ab}
\chi^b_\beta - \frac12 \bar\chi^a_\alpha \varepsilon^{\alpha\beta}
(\bar{\cal Z}R)_{bc} (\Sigma^{bc})_\beta{}^\gamma \chi^a_\gamma +
\nonumber    \\
 && + \bar\Theta_i^M \sqrt\Delta Y_m^{\dot A} (T^M)^{\dot A\dot B}
y^{ma} (\bar\tau^a)^{\alpha i} \Lambda_\alpha^{\dot B} +
\bar\Theta_i^M Z^{m\ddot A}(T^M)^{\ddot A \ddot B} y_{ma}
(\bar\tau^a)^{\alpha i} \Sigma_\alpha^{\ddot B} + \nonumber  \\
 && + \frac14 \bar\Sigma_\alpha (\bar{\cal Z} R)_{ab}
(\Sigma^{ab})_\beta{}^\alpha \varepsilon^{\beta\gamma} \Sigma_\gamma
- \frac14 \bar\Lambda_\alpha(\bar{\cal Z} R)_{ab}
(\Sigma^{ab})_\beta{}^\alpha \varepsilon^{\beta\gamma} \Lambda _
\gamma   -      \nonumber   \\
 && - \left. \frac12 f^{MNK} \left( \frac{i}2 \bar{\Psi}_{\mu i}
\gamma^\mu {\cal Z}^N \bar{\cal Z}^K \Theta_i^M - \varepsilon^{ij}
\bar{\Theta}_i^M {\cal Z}^N \Theta_j^K + \frac1{\sqrt{2}}
\varepsilon^{ij} \bar{\Theta}_i^M {\cal Z}^N \bar{\cal Z}^K \rho_j
\right) \right\},      \label{71}   \\
 {\cal L'^B} &=& e^{ -\hat\varphi \sqrt 2} \left\{ - 2 (\bar{\cal Z}
R)_{ab} ({\cal Z} R)_{ab} - R_{ab} (R_{ab} + \frac12
\varepsilon_{abcd} R_{cd}) +     \right. \nonumber   \\
 && + \left.  \frac{\Delta}2 g^{mn} Y_m \bar{\cal Z}{\cal Z} Y_n +
\frac12 g_{mn} Z^m \bar{\cal Z} {\cal Z} Z^n + \frac18
e^{-\hat\varphi \sqrt{2}} (f^{MNK} {\cal Z}^N \bar{\cal Z}^K)^2
\right\}       \label{72}
\end{eqnarray}
and to the supertransformation laws, respectively,
\begin{eqnarray}
 \delta' \Psi_{\mu i} &=& \frac{i}2 e^{-\hat\varphi/\sqrt 2}
\gamma_\mu \varepsilon^{ij} ({\cal Z} R)_{ab} (\Sigma^{ab})_j{}^k
\eta_k, \nonumber   \\
 \delta' \chi^a_\alpha &=& - 2 e^{-\hat\varphi/\sqrt 2} ({\cal Z}
R)_{ab} (\tau^b)_{i\alpha} \varepsilon^{ij} \eta_j,  \nonumber   \\
 \delta' \rho_i &=& - \frac1{\sqrt 2} e^{-\hat\varphi/\sqrt 2} 
({\cal Z} R)_{ab} (\Sigma^{ab})_i{}^j \eta_j,       \nonumber   \\
 \delta' \Lambda_\alpha^{\dot A} &=& - e^{-\hat\varphi/\sqrt 2}
\varepsilon_{\alpha\beta} \sqrt\Delta {\cal Z}^{\dot A\dot B}
Y_m^{\dot B} y^{ma} (\bar\tau^a)^{\beta i} \eta_i,     \nonumber   \\
 \delta' \Sigma_\alpha^{\ddot A} &=& - e^{-\hat\varphi/\sqrt 2}
\varepsilon_{\alpha\beta} {\cal Z}^{\ddot A\ddot B} Z^{m\ddot B}
y_{ma} (\bar\tau^a)^{\beta i}\eta_i,  \nonumber   \\
 \delta' \Theta_i &=& e^{-\hat\varphi/\sqrt 2} \{ R + \frac12 {\cal
Z} (\bar{\cal Z} R) \frac12 \bar{\cal Z} ({\cal Z} R) \}_{ab}
(\Sigma^{ab})_i{}^j \eta_j + \nonumber  \\
 && + \frac12 e^{-\hat\varphi/\sqrt{2}} f^{MNK} {\cal Z}^N 
\bar{\cal Z}^K \eta_i,                \label{73}
\end{eqnarray}
where
\begin{equation}
 R^M_{ab} = y_{ma} \{ \frac12 (M^M)^{mn} + \frac12 Z^m T^M Z^n +
\frac14 \varepsilon^{mnpq} Y_p T^M Y_q \} y_{nb}.    \label{74}
\end{equation}

Now one can investigate minimum of the potential $V=-{\cal L'^B}$,
defined above. Without losing the generality one can always choose:
\begin{equation}
<y_{ma}>=\delta_{ma}, \qquad <{\cal Z}^M>=(1,i,0,...,0) \label{75}
\end{equation}
In this case the potential has the minimum at $<Y_m> = <Z^m> = 0$ and
one can easily calculate the value of the potential at the minimum:
\begin{equation}
 V_0 = \frac14 M^M_{ab} \{ M^M_{ab} - \frac12 \varepsilon_{abcd}
M^M_{cd}\},   \label{76}
\end{equation}
where $M=1,2$. If the matrices $M^M_{ab}$ are self-dual, then we have
the spontaneous supersymmetry breaking and the cosmological constant
vanishes as a result of the gauge group choice (i. e., which global
translations were made to be local ones) and not of a fine tuning of
the parameters. Let us choose $M^1_{12}=M^1_{34}=m_1$,
$M^2_{14}=M^2_{23}=m_2$ and the other parameters equal zero. In this
case diagonalized gravitino mass matrix has the form:
\begin{equation}
 M^{ik} = \varepsilon^{ij} < ({\cal Z} R)_{ab} > (\Sigma^{ab})_j{}^k
\sim \pmatrix{m_1 + m_2 & 0 \cr 0 & m_1 -m_2}.      \label{77}
\end{equation}

Unfortunately, the spontaneous symmetry breaking does not generate
the masses for the Y- and Z-sectors. It is easy to check, that
matrices $(\Sigma^{ab})_\alpha{}^\beta$ are anti-self-dual and
spinors $\Lambda_\alpha$ and $\Sigma_\alpha$ do not acquire masses
because of a vanishing of the expression $<(\bar{\cal Z}R)_{ab}>
(\Sigma^{ab})_\alpha{}^\beta$. Scalar fields $Y_m$ and $Z^m$ also
remain massless, which can be seen taking into account that the
generators $(M^M)^{mn}$, $(T^M)^{\dot A\dot B}$ and $(T^M)^{\ddot
A\ddot B}$ are nontrivial only when $M=1,2$, $M=3,...,3+p$ and
$M=4+p,...,4+p+q$, correspondingly.

\section{V(p,q)-model}

The N=2 supergravity model with the second general type of
nonsymmetric quaternionic geometry, a so called V(p,q)-model
\cite{Ale75,Cec89}, has also been constructed in \cite{Tso96}. Here
we again refrain from a detailed description of it, paying the main
attention to the symmetry breaking in this model.

\subsection{Description of the model}

The hidden sector of this model is essentially the same as in the
previous one, but in different parameterization. It contains the
following fields: graviton $e_{\mu r}$, gravitini $\Psi_{\mu i}$,
fermions $\lambda_a^i$ and $\chi^i$, where $i=1,2$ and $a=1,2,3$, and
bosonic fields $y_{ma}$, $\varphi$, $\pi^{[mn]}$, $l^m$ and $\pi_m$,
where $m=1,2,3$. There are three sets of hypermultiplets interacting
with the hidden sector:  $(\Omega^i, X^m, Z)^A$, $(\Lambda^i,
Y_m,Y)^{\dot A}$ and $(\Sigma^i,Z^m,Z)^{\ddot A}$, which we will call
$X$-, $Y$- and $Z$-multiplets, correspondingly, and it turns out to
be necessary to introduce $\gamma$-matrices $\Gamma^{A\dot A\ddot A}$
in order to connect fields from different kinds of the multiplets in
the Lagrangian. The $X$-multiplet carries vector index $A$ of the
$O(p)$ group and $Y$- and $Z$-multiplets carry the corresponding
spinor indices in full correspondence with \cite{Ale75,Cec89}. 

The fermionic Lagrangian of the V(p,q)-model has the following form:
\begin{eqnarray}
 \cal{L}^F &=& \frac{i}2 \varepsilon^{\mu\nu\rho\sigma}
\bar{\Psi}_{\mu i} \gamma_5 \gamma_{\nu} D_{\rho} \Psi_{\sigma i} +
\frac{i}2 \bar{\lambda}_a^i \hat{D} \lambda_a^i + \frac{i}2
\bar{\chi}^i \hat{D} \chi^i +  \nonumber  \\
 && + \frac{i}2 \bar{\Omega}^i \hat{D} \Omega^i + \frac{i}2
\bar{\Lambda}^i \hat{D} \Lambda^i + \frac{i}2 \bar{\Sigma}^i \hat{D}
\Sigma^i -   \nonumber   \\
 && - \frac12 \bar{\chi}^i \gamma^{\mu} \gamma^{\nu} \left\{
\partial_{\nu} \varphi \delta_i{}^j - 2 e^{\varphi} Q^{a+}_{\nu}
\tau^a{}_i{}^j \right\} \Psi_{\mu j} -   \nonumber \\
 && - \frac12 \bar{\lambda}^i_a \gamma^{\mu} \gamma^{\nu} \left\{
(S^+_{\nu} + P_{\nu})_{ab} \tau^b{}_i{}^j + 2 e^{\varphi}
Q^{a-}_{\nu} \delta_i{}^j \right\} \Psi_{\mu j} -   \nonumber  \\
 && - \frac12 \bar{\Psi}_{\mu i} \gamma^{\nu} \gamma^{\mu} \left\{
e^{\varphi} D_{\nu} X\delta_i{}^j + y_{ma} \partial_{\nu} X^m
\tau^a{}_j{}^i \right\} \Omega^j - \nonumber   \\
 && - \frac12 e^{\varphi/2} \bar{\Psi}_{\mu i} \gamma^{\nu}
\gamma^{\mu} (V_{\nu})_i{}^j \Lambda^j - \frac12 e^{\varphi/2}
\bar{\Psi}_{\mu i} \gamma^{\nu} \gamma^{\mu} (W_{\nu})_i{}^j \Sigma^j
+   \nonumber  \\
 && + \frac{i}2 (S^-_{\mu} + P_{\mu})_{ab} (\bar{\lambda}^i_a
\gamma_{\mu} \lambda^i_b) + 2 i e^{\varphi} Q^{a-}_{\mu}
(\bar{\lambda}^i_a \gamma_{\mu} \chi^i) - \nonumber  \\
 && - (\bar{\lambda}^i_a \gamma^{\mu} \Omega^i) y_{ma} \partial_{\mu}
X^m - i (\bar{\chi}^i \gamma^{\mu} \Omega^i) e^{\varphi} D_{\mu} X -
\nonumber \\
 && - \frac{i}2 e^{\varphi/2} \bar{\chi}^i \gamma^{\mu}
(V_{\mu})_i{}^j \Lambda^j - \frac{i}2 e^{\varphi/2} \bar{\chi}^i
\gamma^{\mu} (W_{\mu})_i{}^j \Sigma^j -  \nonumber   \\
 && - \frac{i}2 e^{\varphi/2} \bar{\lambda}^i_a \gamma^{\mu}
\tau^a{}_j{}^i (V_{\mu})_k{}^j \Lambda^k + \frac{i}2 e^{\varphi/2}
\bar{\lambda}^i_a \gamma^{\mu} \tau^a{}_j{}^i (W_{\mu})_k{}^j
\Sigma^k -   \nonumber    \\ 
 && - \frac{i}2 e^{\varphi/2} \Gamma^{A\dot{A}\ddot{A}}
\bar{\Omega}^{iA} \gamma^{\mu} (V_{\mu})_j{}^i \Sigma^{j\ddot{A}} + 
\frac{i}2 e^{\varphi/2} \Gamma^{A\dot{A}\ddot{A}} \bar{\Omega}^{iA}
\gamma^{\mu} (W_{\mu})_j{}^i \Lambda^{j\dot{A}} -   \nonumber   \\
 && - \frac{i}2 \Gamma^{A\dot{A}\ddot{A}} \bar{\Lambda}^{i\dot{A}}
\gamma^{\mu} \left\{ e^{\varphi} D_{\mu} X^A \delta_j{}^i -
\partial_{\mu} X^{mA} y_{ma} \tau^a{}_j{}^i \right\}
\Sigma^{j\ddot{A}}.    \label{1}
\end{eqnarray}

The conventions in this Lagrangian are the following:
\begin{eqnarray}
 P_{\mu ab} &=& y_{ma} \left\{ \partial_{\mu} \pi^{mn} + \frac12
(X^m \stackrel{\leftrightarrow}{\partial_{\mu}} X^n) \right\} y_{nb},
\nonumber   \\
 S^{\pm}_{\mu ab} &=& \frac12 [y^{ma} \partial_{\mu} y_{mb} \pm
y^{mb} \partial_{\mu} y_{ma}],    \quad Q^{a\pm}_{\mu} = y_{ma}
L^m_{\mu} \pm \frac14 y^{ma} U_{\mu m}, \nonumber   \\
 U_{\mu m} &=& \partial_\mu \pi_m + (Y_m^{\dot{A}}
\stackrel{\leftrightarrow}{\partial_{\mu}} Y^{\dot A}) - \frac12
\varepsilon_{mnk} (Z^{n\ddot A} \stackrel {\leftrightarrow}
{\partial_{\mu}} Z^{k\ddot A}),      \nonumber  \\
 D_{\mu}X^A &=& \partial_{\mu} X^A + X^{mA} U_{\mu m} + \frac12
\Gamma^{A\dot{A} \ddot{A}} [(Y^{\dot{A}} \stackrel{\leftrightarrow}
{\partial_{\mu}} Z^{\ddot{A}}) + (Y_m^{\dot{A}}
\stackrel{\leftrightarrow}{\partial_{\mu}} Z^{m\ddot{A}})], \nonumber
\\
 L_{\mu}^m &=& \partial_{\mu} l^m + \frac12 \pi^{mn} U_{\mu n} +
\frac12 X^{mA} D_{\mu} X^A - \frac14 X^{mA} X^{nA} U_{\mu n} -
\nonumber  \\
 && - \frac18 \varepsilon^{mnk} (Y_n^{\dot{A}}
\stackrel{\leftrightarrow}{\partial_{\mu}} Y_k^{\dot{A}}) + \frac14
(Z^{m\ddot{A}} \stackrel{\leftrightarrow} {\partial_{\mu}}
Z^{\ddot{A}}),    \nonumber  \\
 (V_{\mu})_i{}^j &=& \frac1{\sqrt{\Delta}} \partial_{\nu} Y
\delta_i{}^j + \sqrt{\Delta} D_{\nu} Y_my^{ma} \tau^a{}_i{}^j,
\nonumber  \\
 (W_{\mu})_i{}^j &=& \sqrt{\Delta} D_{\mu} Z \delta_i{}^j +
\frac1{\sqrt{\Delta}} D_{\mu} Z^my_{ma} \tau^a{}_i{}^j, \nonumber \\
 D_{\mu} Z^{m\ddot{A}} &=& \partial_{\mu} Z^{m\ddot{A}} +
\Gamma^{A\dot{A}\ddot{A}} X^{mA} \partial_{\mu} Y^{\dot A}, \nonumber
\\
 D_{\mu} Y_m^{\dot A} &=& \partial_{\mu} Y_m^{\dot A} -
\varepsilon_{mnk} [\pi^{nk} \delta^{\dot{A}\dot{B}} + \frac12 X^{nA}
X^{kB} (\Sigma^{AB})^{\dot{A}\dot{B}}] \partial_{\mu} Y^{\dot{B}} -
\nonumber   \\
 && - \varepsilon_{mnk} X^{nA} \Gamma^{A\dot{A}\ddot{A}}
\partial_{\mu} Z^{k\ddot{A}}, \nonumber   \\
 D_{\mu} Z^{\ddot{A}} &=& \partial_{\mu} Z^{\ddot{A}} +
\varepsilon_{mnk} [\pi^{mn} \delta^{\ddot{A}\ddot{B}} + \frac12
X^{mA} X^{nB} (\Sigma^{AB})^{\ddot{A} \ddot{B}}] \partial_{\mu}
Z^{k\ddot{B}} -    \nonumber   \\
 && - X^{mA} \Gamma^{A\dot{A}\ddot{A}} (\partial_{\mu} Y_m^{\dot{A}}
- \varepsilon_{mnk} \pi^{nk} \partial_{\mu} Y^{\dot{A}}) - \nonumber
\\
 && - \frac16 \varepsilon_{mnk} X^{mA} X^{nB} X^{kC}
(\Gamma^{ABC})^{\dot{A}\ddot{A}} \partial_{\mu}Y^{\dot{A}}  \label{2}
\end{eqnarray}
and D-derivatives for the fermions have the following form:
\begin{equation}
 (D_{\mu})_i{}^j = D_{\mu}^G \delta_i^j \pm \frac14 \varepsilon^{abc}
(S_{\mu}^- - P_{\mu})_{ab} \tau^c{}_i{}^j + e^{\varphi} Q^{a+}_{\mu}
\tau^a{}_i{}^j. \label{3}
\end{equation}
Here derivatives of $\Psi_{\mu i}$ and $\eta_i$ have the sign "-" and 
derivatives of all the other fermion fields -- the sign "+".

The corresponding supertransformations of the fermionic fields are
the following:
\begin{eqnarray}
 \delta \Psi_{\mu i} &=& 2 D_{\mu} \eta_i \qquad \delta \chi^i = - i
\gamma_{\mu} \left\{ \partial_{\mu} \varphi \delta_i{}^j - 2
e^{\varphi} Q^{a+}_{\mu} \tau^a{}_i{}^j \right\} \eta_j, \nonumber
\\
 \delta \lambda^i_a &=& - i \gamma^{\mu} \left\{ (S^+_{\mu} +
P_{\mu})_{ab} \tau^b{}_i{}^j + 2 e^{\varphi} Q^{a-}_{\mu}
\delta_i{}^j \right\} \eta_j \nonumber,    \\
 \delta \Omega^i &=& - i \gamma^{\mu} \left\{ e^{\varphi} D_{\mu} X
\delta_i^j + y_{ma} \partial_{\mu} X^m \tau^a{}_i{}^j \right\}
\eta_j, \nonumber     \\
 \delta \Lambda^i &=& - i \gamma^{\mu} (V_{\mu})_i{}^j \eta_j,
\qquad \delta \Sigma^i = - i \gamma^{\mu} (W_{\mu})_i{}^j \eta_j.
\label{4}
\end{eqnarray}
Here we use the same conventions as in (\ref{2}) and (\ref{3}).

The bosonic Lagrangian of the V(p,q)-model has the following form:
\begin{eqnarray}
 \cal{L}^B &=& \frac12 (\partial_{\mu} \varphi)^2 + \frac12 (S_{\mu
ab}^+)^2 + \frac12 (P_{\mu ab})^2 + 4 e^{2\varphi} g_{mn} L_{\mu}^n
L_{\nu}^n + \nonumber  \\
 && + \frac14 e^{2\varphi} g^{mn} U_{\mu m} U_{\mu n} + \frac12
e^{2\varphi} (D_{\mu}X)^2 + \frac12 g_{mn} \partial_{\mu} X^m
\partial_{\mu} X^n + \frac{e^{\varphi}}{2\Delta} (\partial_{\mu} Y)^2
+ \nonumber   \\
 && + e^{\varphi} \frac{\Delta}2 (D_{\mu} Y_m) (D_{\mu} Y_n) g^{mn} +
e^{\varphi} \frac{\Delta}2 (D_{\mu} Z)^2 +
\frac{e^{\varphi}}{2\Delta} (D_{\mu}Z^m) (D_{\mu} Z^n) g_{mn},
\label{5}
\end{eqnarray}
where $g_{mn}=y_{ma}y_{na}$ and $g^{mn}=y^{ma}y^{na}$. We do not give
here the corresponding supertransformations of the bosonic fields,
because they are awkward and nonessential for our considerations.

Now let us consider interaction of the V(p,q)-model with the vector
multiplets, described in Section I. It is easy to check, that the
only additional terms, which appear in the fermionic Lagrangian, are
the following:
\begin{eqnarray}
 \Delta {\cal L^F} &=& \frac{e^{\hat\varphi/\sqrt2}}8
\varepsilon_{ij} \left\{ (\bar\chi^i \bar{\cal Z}^M (\sigma A)^M
\chi^j) + (\bar\lambda^i_a \bar{\cal Z}^M (\sigma A)^M \lambda^j_a) +
\right.  \nonumber     \\
 && + \left. (\bar\Omega^i \bar{\cal Z}^M (\sigma A)^M \Omega^j) +
(\bar\Lambda^i \bar{\cal Z}^M (\sigma A)^M \Lambda^j) + (\bar\Sigma^i
\bar{\cal Z}^M (\sigma A)^M\Sigma^j) \right\}.      \label{17}
\end{eqnarray}
Besides, D-derivatives of the fermionic fields change their form.
Derivatives (\ref{3}) of all the fermions of the V(p,q)-model acquire
the following additional terms (analogously to the ones in
(\ref{13})):
\begin{equation}
 \frac14 (\bar{\cal Z} \partial_\mu {\cal Z}) - \frac1{2\sqrt 2}
e^{-\sqrt 2 \hat\varphi} \gamma_5 \partial_\mu \hat\pi   \label{18}
\end{equation}
and derivatives (\ref{13}) of the fields $\rho_i$ and $\Theta_i^M$
acquire the following additional terms (analogously to the ones in
(\ref{3})):
\begin{equation}
 \frac14 \varepsilon^{abc} (S_{\mu}^- - P_{\mu})_{ab} \tau^c{}_i{}^j
+ e^{\varphi} Q^{a+}_{\mu} \tau^a{}_i{}^j. \label{19}
\end{equation}

\subsection{Gauge interaction and symmetry breaking}

To investigate the possibilities of the symmetry breaking, we have to
switch on a gauge interaction in the hidden sector of the model. The
hidden sector is the same as for $W(p,q)$-model, it has just been
rewritten in other variables. And it has translations
$\pi_m\to\pi_m+\Lambda_m$ and $l^m\to l^m+\Lambda^m$ as part of its
global symmetry group. These translations correspond to the
translation of the field $\pi^{mn}$ in $W(p,q)$-model. And, as it has
been shown in the previous section, by making part of these
translations local one can obtain the spontaneous symmetry breaking
with two arbitrary mass scales and vanishing cosmological constant.

In order to learn, if the mass splitting in the $X$-, $Y$- and
$Z$-multiplets appear in our model, we also have to switch on the
gauge interaction, which touches the corresponding sectors. The
question, we are interested to answer as well, is: if the Yukawa
couplings, mixing the fields from the different multiplets, appear in
the Lagrangian after the symmetry breaking.

In general case there are two global symmetries of the sector,
including $X$-,$Y$- and $Z$-multiplets, which do not touch the hidden
sector. The first one is the following:
\begin{equation}
 \delta X^A = \alpha^{M_1} (T_1^{M_1})^{AB} X^B, \quad \delta
 Y^{\dot A} = \alpha^{M_1} (\dot T_1^{M_1})^{\dot A\dot B} Y^{\dot
B}, \quad \delta Z^{\ddot A} = \alpha^{M_1} (\ddot T_1^{M_1})^{\ddot
A\ddot B}Z^{\ddot B}, \label{40}
\end{equation}
where generators $\dot{T}_1$ and $\ddot{T}_1$ are connected to
generator $T_1$:
\begin{equation}
 (\dot T_1^{M_1})^{\dot A\dot B} = \frac14 (T_1^{M_1})^{AB}
(\Sigma^{AB})^{\dot A\dot B},   \qquad (\ddot T_1^{M_1})^{\ddot
A\ddot B} = \frac14 (T_1^{M_1})^{AB} (\Sigma^{AB})^{\ddot A\ddot B}.
\label{41}
\end{equation}
The generators $T_1$ are chosen to be real, antisymmetric and
correspond to some subgroup of the orthogonal group $O(p)$, where $p$
is the number of the $X$-multiplets. The fields of the $X$-multiplets
transform under the vector representation of this group and the
fields of the $Y$- and $Z$-multiplets transform under the spinor
representation.

As it has been shown in \cite{Wit93,Wit94}, depending of the values
of $p,q$, there exists an additional global symmetry group:
\begin{equation}
 \delta Y^{\dot A} = \alpha^{M_2} (\dot T_2^{M_2})^{\dot A\dot B}
Y^{\dot B}, \qquad \delta Z^{\ddot A} = \alpha^{M_2} (\ddot
T_2^{M_2})^{\ddot A\ddot B} Z^{\ddot B}, \qquad   \delta X^A = 0,
\label{42}
\end{equation}
where generators $\dot{T}_2$ and $\ddot{T}_2$ commute with
$\Gamma$-matrices:
\begin{equation}
 \dot T_2^{\dot A\dot B} (\Gamma^A)^{\dot B\ddot A} -
(\Gamma^A)^{\dot A\ddot B} \ddot T_2^{\ddot B\ddot A} = 0.
\label{43}
\end{equation}
The generators $T_i$ have to obey the following commutation
relations:
\begin{equation}
 T_i^{M_i} \cdot T_i^{N_i} - (M_i \leftrightarrow N_i) =
f_i^{M_iN_iK_i} T_i^{K_i}, \label{44}
\end{equation}
where $f_i^{M_iN_iK_i}$ are the structure constants of the
corresponding symmetry groups and the indices $M_i$ are the indices
of the adjoint representations. 

Let us denote these two sets of indices by a general index  $M$:
$M=\{(M_1),(M_2)\}$, structure constants of the direct product
of these symmetry groups by $f^{MNK}$ and
\begin{eqnarray}
 (\dot T^M)^{\dot A\dot B} &=& \{ (\dot T_1^{M_1})^{\dot A\dot B},
(\dot T_2^{M_2})^{\dot A\dot B}\},   \qquad (\ddot T^M)^{\ddot A\ddot
B} = \{ (\ddot T_1^{M_1})^{\ddot A\ddot B}, (\ddot T_2^{M_2})^{\ddot
A\ddot B}\},    \nonumber    \\
 (T^M)^{AB} &=& \{ (T_1^{M_1})^{AB}, 0\}.      \label{45}
\end{eqnarray}
Then the commutation relations (\ref{44}) can be rewritten in the
general form:
\begin{equation}
 T^{M} \cdot T^{N} - (M \leftrightarrow N) = f^{MNK} T^{K} \label{46}
\end{equation}
and relationship (\ref{43}), taking into account that $\Gamma^A
\Sigma^{BC} - \Sigma^{BC} \Gamma^A = 2 (\delta^{AB} \Gamma^C -
\delta^{AC} \Gamma^B)$, can be rewritten in the following form:
\begin{equation}
 (\dot T^M)^{\dot A\dot B} (\Gamma^A)^{\dot B\ddot A} -
(\Gamma^A)^{\dot A\ddot B} (\ddot T^M)^{\ddot B\ddot A} = (T^M)^{AB}
(\Gamma^B)^{\dot A\ddot A}.   \label{47}
\end{equation}

Now we can switch on the gauge interaction. For that let us make the 
following substitution in the Lagrangian and the supertransformation
laws:
\begin{eqnarray}
 \partial_\mu {\cal Z}^M &\to& \partial_\mu {\cal Z}^M - f^{MNK}
A_\mu{}^N {\cal Z}^K, \qquad \partial_\mu \Theta_i{}^M \to
\partial_\mu \Theta_i{}^M - f^{MNK} A_\mu{}^N \Theta_i{}^K, \nonumber
\\
 \partial_\mu l^m &\to& \partial_\mu l^m - A_\mu^M M^{Mm}, \qquad 
\partial_\mu \pi_m \to \partial_\mu \pi_m - A_\mu^M N^M_m,
\label{48} \\
 \partial_\mu X^A &\to& \partial_\mu X^A - A_\mu^M (T^M)^{AB} X^B,
\qquad \partial_\mu \Omega^A \to \partial_\mu \Omega^A - A_\mu^A
(T^M)^{AB} \Omega^B \nonumber
\end{eqnarray}
and analogous expressions for the fields from $Y$- and
$Z$-multiplets.

In order to restore the invariance of the Lagrangian under the
supertransformations, one has to add to the Lagrangian the following
terms:
\begin{eqnarray}
 {\cal L'^F} &=& e^{-\hat\varphi/\sqrt 2} \left\{ -\frac14
\bar\Psi_{\mu i} \sigma^{\mu\nu} \varepsilon^{ij} (R_+^a + G^a)
\tau^a{}_j{}^k \Psi_{\nu k} +     \right.      \nonumber   \\
 && + \frac{i}{2\sqrt 2} \bar\Psi_{\mu i} \gamma^\mu (\bar R_+^a +
\bar G^a) \tau^a{}_i{}^j \rho_j - \frac{i}2 \bar\Psi_{\mu i}
\gamma^\mu (R_+^{Ma} + G^{Ma}) \tau^a{}_i{}^j \Theta^M_j - \nonumber
\\
 && - \frac{i}2 \bar\Psi_{\mu i} \gamma^\mu \bar R_+^a \tau^a{}_i{}^j
\varepsilon_{jk} \chi^k - \frac{i}2 \bar\Psi_{\mu i} \gamma^\mu
\varepsilon_{ij} (\bar R_-^a \delta_j{}^k - \varepsilon^{abc} \bar 
G^b \tau^c{}_j{}^k) \lambda^k_a + \nonumber  \\
 && + \frac1{\sqrt 2} \bar\rho_i \bar R_+^a \tau^a{}_j{}^i \chi^j -
\frac1{\sqrt 2} \bar\rho_i (\bar R_-^a \delta_i{}^j -
\varepsilon^{abc} \bar G^b \tau^c{}_i{}^j) \lambda_a^j - \bar\chi^i
\varepsilon_{ij} \bar R_-^a \lambda^j_a + \nonumber  \\
 && + \frac1{\sqrt 2} \bar\rho_i \varepsilon^{ij} (R_+^{Ma} + G^{Ma})
\tau^a{}_j{}^k \Theta_k^M + \frac14 \bar\lambda^i_a \varepsilon_{ij}
(\bar R_+^b - \bar G^b) \tau^b{}_k{}^j \lambda^k_a +  \nonumber   \\
 && + \bar\lambda^i_a (R_-^{Ma} \delta_i{}^j - \varepsilon^{abc}
G^{Mb}\tau^c{}_i{}^j) \Theta_j^M - \frac14 \bar\Theta^M_i
\varepsilon_{ij} (R_+^a - G^a) \tau^a {}_j{}^k \Theta^M_k + \nonumber
\\
 && + \frac14 \bar\chi^i \varepsilon_{ij} (\bar R_+^a - \bar G^a)
\tau^a{}_k{}^j \chi^k - \bar\chi^i R_+^{Ma} \tau^a{}_i{}^j \Theta^M_j
- \frac12 \varepsilon^{abc} \bar\lambda^i_a \varepsilon_{ij}\bar G^c
\lambda^j_b + \nonumber        \\
 && + \frac{i}2 \bar\Omega^{iA} \gamma^\mu \varepsilon^{ij} ({\cal
F}_X{}^A)_j{}^k \Psi_{\mu k} + \frac{i}2 \bar\Lambda^{i\dot A}
\gamma^\mu \varepsilon^{ij} ({\cal F}_Y{}^{\dot A})_j{}^k \Psi_{\mu
k} + \frac{i}2 \bar\Sigma^{i\ddot A} \gamma^\mu \varepsilon^{ij}
({\cal F}_Z{}^{\ddot A})_j{}^k \Psi_{\mu k} \nonumber  \\
 && + \frac1{\sqrt 2} \bar\Omega^{iA} (\bar{\cal
F}_X{}^A)_i{}^j\rho_j + \frac1{\sqrt 2} \bar\Lambda^{i\dot A}
(\bar{\cal F}_Y{}^{\dot A})_i{}^j \rho_j + \frac1{\sqrt 2}
\bar\Sigma^{i\ddot A} (\bar{\cal F}_Z{}^{\ddot A})_i{}^j \rho_j -
\nonumber  \\
 && - \bar\Omega^{iA} ({\cal F}_X^{MA})_i{}^j \Theta^M_j -
\bar\Lambda^{i\dot A} ({\cal F}_Y^{M\dot A})_i{}^j \Theta_j^M -
\bar\Sigma^{i\ddot A} ({\cal F}_Z^{M\ddot A})_i{}^j \Theta^M_j -
\nonumber  \\
 && + \frac12 \bar\Lambda^{i\dot A} (\bar{\cal F}_Y{}^{\dot A})_i{}^j
\varepsilon_{jk} \chi^k - \frac12 \bar\Lambda^{i\dot A}
\tau^a{}_i{}^j (\bar{\cal F}_Y{}^{\dot A})_j{}^k \varepsilon_{kp}
\lambda^p_a - e^{\varphi} \bar\Omega^{iA} \varepsilon_{ij} \bar
F_X{}^A\chi^j + \nonumber    \\
 && + \frac12 \bar\Sigma^{i\ddot A} (\bar{\cal F}_Z{}^{\ddot
A})_i{}^j \varepsilon_{jk} \chi^k + \frac12 \bar\Sigma^{i\ddot A}
\tau^a{}_i{}^j (\bar{\cal F}_Z{}^{\ddot A})_j{}^k \varepsilon_{kp}
\lambda^p_a - \bar\Omega^{iA} \varepsilon_{ij} \bar{\cal Z}^{AB}
X^{mB} y_{ma} \lambda^j_a - \nonumber    \\
 && - \frac12 \Gamma^{A\dot A\ddot A} \{\bar\Sigma^{i\ddot A}
\varepsilon_{ij} (\bar{\cal F}_X{}^A)_k{}^j \Lambda^{k\ddot A} +
\bar\Lambda^{i\dot A}(\bar {\cal F}_Z{}^{\ddot A})_i{}^j
\varepsilon_{jk} \Omega^{kA} - \bar\Sigma^{j\ddot A} (\bar{\cal
F}_Y{}^{\ddot A})_i^j \varepsilon_{jk} \Omega^{kA}\} - \nonumber   \\
 && - \frac12 \bar\Omega^{iA} \varepsilon_{ij} \bar{\cal Z}^{AB}
\Omega^{jB} - \frac12 \bar\Lambda^{i\dot A} \varepsilon_{ij}
\bar{\cal Z}^{\dot A\dot B} \Lambda^{j\dot B} - \frac12
\bar\Sigma^{i\ddot A} \varepsilon_{ij} \bar{\cal Z}^{\ddot A\ddot B}
\Sigma^{j\ddot B} +           \nonumber  \\
 && + \frac14 (R_+^{Ma} - G^{Ma}) (\bar\Omega^i \varepsilon_{ij}
\bar{\cal Z}^M \tau ^a{}_k{}^j \Omega^k) + \nonumber \\
 && + \frac14 (R_-^{Ma} -
G^{Ma})  \{ \bar\Lambda^i \varepsilon_{ij} \bar{\cal Z}^M
\tau^a{}_k{}^j \Lambda^k + \bar\Sigma^i \varepsilon_{ij} \bar{\cal
Z}^M \tau^a{}_k{}^j \Sigma^k\} - \nonumber  \\
 && - \left. \frac12 f^{MNK} \left( \frac{i}2 \bar{\Psi}_{\mu i}
\gamma^\mu {\cal Z}^N \bar{\cal Z}^K \Theta_i^M - \varepsilon^{ij}
\bar{\Theta}_i^M {\cal Z}^N \Theta_j^K + \frac1{\sqrt{2}}
\varepsilon^{ij} \bar{\Theta}_i^M{\cal Z} ^N \bar{\cal Z}^K \rho_j
\right)\right\},           \label{49}   \\
 {\cal L'^B} &=& - \frac12 e^{-\sqrt 2 \hat\varphi} \left\{ (R_+^{Ma}
+ G^{Ma})^2 + |R_+^a|^2 + |R_-^a|^2 + 2|G^a|^2 + e^{2\varphi}
|F_X{}^A|^2 \right. - \nonumber  \\
 && - g_{mn} (X^m{\cal Z} \bar{\cal Z} X^n) -
\frac{e^{\varphi}}{\Delta} (Y{\cal Z} \bar {\cal Z}Y) + \Delta
e^{\varphi} g^{mn} F_Y{}_m^{\dot A} \bar F_Y{}_n^{\dot A}+ \Delta
e^{\varphi} |F_Z{}^{\ddot A}|^2 + \nonumber  \\
 && + \left. \frac{e^{\varphi}}{\Delta} g_{mn} F_Z{}^{m\ddot A} \bar
F_Z{}^{n\ddot A} + \frac18 (f^{MNK} {\cal Z}^N \bar{\cal Z}^K)^2
\right\}, \label{50}
\end{eqnarray}
where the following notations are used:
\begin{eqnarray}
 R_\pm^{Ma} &=& e^{\varphi} \{ y_{ma} (A^{mM} + 2 \pi^{mn} B_n^M +
X^m X^n B_n^M) \pm y^{ma} B_m^M\},     \nonumber   \\
 A^{mM} &=& - 2 M^{mM} + X T^M X^m - \frac12 X^{mA} \{ Y \Gamma^A
{\ddot T}^M Z + Y {\dot T}^M \Gamma^A Z +     \nonumber   \\
 && + Y_m \Gamma^A {\ddot T}^M Z^m + Y_m {\dot T}^M \Gamma^A Z^m\} +
\frac12 \varepsilon^{mnk} Y_n {\dot T}^M Y_k + Z {\ddot T}^M Z^m,
\nonumber  \\
 B_m^M &=& - \frac12 N_m^M + Y{\dot T}^M Y_m + \frac12
\varepsilon_{mnk} Z^n {\ddot T}^M Z^k,          \nonumber  \\
 G^M_a &=& \frac12 \varepsilon^{abc} y_{mb} (X^m T^M X^n) y_{nc},
\nonumber \\
 ({\cal F}_X^{MA})_i{}^j &=& e^{\varphi} F_X^{MA} \delta_i{}^j +
(T^M)^{AB} X^{mB} y_{ma} \tau^a{}_i{}^j,            \nonumber \\
 ({\cal F}_Y^{M\dot A})_i{}^j &=& e^{\varphi/2} \{ \frac1
{\sqrt\Delta} (T^M)^{\dot A\dot B} Y^{\dot B} \delta_i{}^j +
\sqrt\Delta F_{Ym}^{M\dot A} y^{ma} \tau^a{}_i{}^j \}, \nonumber  \\
 ({\cal F}_Z^{M\ddot A})_i{}^j &=& e^{\varphi/2} \{ \sqrt\Delta
F_Z^{M\ddot A} \delta_i{}^j + \frac1{\sqrt\Delta} F_Z^{Mm\ddot A}
y_{ma} \tau^a{}_i{}^j \}, \nonumber  \\
 F_X^{MA} &=& (T^M)^{AB} X^B - 2 X^{mA} B_m^M + \frac12 X^{mA} \{ Y
\Gamma^A {\ddot T^M} Z + Y {\dot T^M} \Gamma^A Z +   \nonumber  \\
 && + Y_m \Gamma^A {\ddot T^M} Z^m + Y_m {\dot T^M} \Gamma^A Z^m\},
\nonumber  \\
 F_Z^{Mm\ddot A} &=& (T^M)^{\ddot A\ddot B} Z^{m\ddot B} + X^{mA}
(\Gamma^A)^{\dot A\ddot A} (T^M)^{\dot A\dot B} Y^{\dot B}, \nonumber
\\
 F_{Ym}^{M\dot A} &=& (T^M)^{\dot A\dot B} Y_m^{\dot B} -
\varepsilon_{mnk} \{ \pi^{nk} \delta^{\dot A \dot B} + \frac12 X^{nA}
X^{kB} (\Sigma^{AB})^{\dot A \dot B}\} (T^M)^{\dot B\dot C} Y^{\dot
C} -         \nonumber  \\
 && - \varepsilon_{mnk} X^{nA} (\Gamma^A)^{\dot A\ddot A}
(T^M)^{\ddot A \ddot B} Z^{k\ddot B},     \nonumber  \\
 F_Z^{M\ddot A} &=& (T^M)^{\ddot A\ddot B} Z^{\ddot B} +
\varepsilon_{mnk} \{ \pi^{mn} \delta^{\ddot A\ddot B} + \frac12
X^{mA} X^{nB} (\Sigma^{AB})^{\ddot A \ddot B}\} (T^M)^{\ddot B\ddot
C} Z^{k\ddot C} -        \nonumber  \\
 && - X^{mA} (\Gamma^A)^{\dot A\ddot A} (T^M)^{\dot A\dot B}
Y_m^{\dot B} + \varepsilon_{mnk} X^{mA} (\Gamma^A)^{\dot A\ddot B} \{
\pi^{nk} \delta^{\ddot A\ddot B} -		\nonumber  \\
 && - \frac16 X^{nB} X^{kC} (\Sigma^{BC})^{\ddot B\ddot A} \}
(T^M)^{\dot A \dot B}Y^{\dot B}.
\end{eqnarray}

Additional terms to the supertransformation laws are the following:
\begin{eqnarray}
 \delta' \Psi_{\mu i} &=& \frac{i}2 e^{-\hat\varphi/\sqrt 2}
\gamma_\mu \varepsilon^{ij} (R_+^a + G^a) \tau^a{}_j{}^k \eta_k,
\nonumber    \\
 \delta' \chi^i &=& - e^{-\hat\varphi/\sqrt 2} \varepsilon^{ij}
(R_+^a + G^a) \tau^a{}_j{}^k \eta_k, \nonumber    \\
 \delta' \lambda^i_a &=& e^{-\hat\varphi/\sqrt 2} \varepsilon^{ij}
(R_-^a \delta_j{}^k - \varepsilon^{abc} G^b \tau^c{}_j{}^k) \eta_k,
\nonumber    \\
 \delta' \rho_i &=& - \frac1{\sqrt 2} e^{-\hat\varphi/\sqrt 2} (R_+^a
+ G^a) \tau^a{}_i{}^j \eta_j, \nonumber    \\
 \delta' \Theta_i^M &=& e^{-\hat\varphi/\sqrt 2} \{ (R_+^{Ma} +
G^{Ma}) + \frac12 \bar{\cal Z}^M (R_+^a + G^a) +        \nonumber  \\
 && + \frac12 {\cal Z}^M (\bar R_+^a + \bar G^a) \} \tau^a{}_i{}^j
\eta_j + \frac12 e^{-\hat\varphi/\sqrt{2}} f^{MNK} {\cal Z}^N \bar
{\cal Z}^K \eta_i, \nonumber    \\
 \delta' \Omega^{iA} &=& - e^{-\hat\varphi/\sqrt 2} \varepsilon^{ij}
({\cal F}_X{}^A)_j{}^k \eta_k,     \nonumber   \\
 \delta' \Lambda^{i\dot A} &=& - e^{-\hat\varphi/\sqrt 2}
\varepsilon^{ij} ({\cal F}_Y{}^{\dot A})_j{}^k \eta_k,     \quad 
\delta' \Sigma^{i\ddot A} = - e^{-\hat\varphi/\sqrt 2}
\varepsilon^{ij} ({\cal F}_Z{}^{\ddot A})_j{}^k \eta_k. \label{51}
\end{eqnarray}

Let us again choose vacuum expectation values of the scalar fields in
the following form: $<y_{ma}>\sim\delta_{ma}$ and
$<{\cal Z}^M>=(1,i,0,...,0)$. In this, the potential has the minimum
at vanishing vacuum expectation values for the fields $X$, $Y$ and
$Z$, its value at the minimum being:
$V_0=2(M^{Ma}-\frac{1}{4}N^M_a)^2$, where indices $M=1,2$, and
$a=1,2,3$. One can see from this formula, that if one makes the
following choice of the gauge group:
\begin{equation}
 M^{11} = \frac14 N_1^1 = m_1  \qquad  M^{22} = \frac14 N_2^2 = m_2,
\end{equation}
all the other parameters being equal to zero, then it is easy to
check, that a cosmological term vanishes.

The gravitino mass matrix takes the form:
\begin{equation}
 M^{ij} = - \frac12 \varepsilon^{ij} <R_+^a> \tau^a{}_j{}^k \sim
\pmatrix{m_1 + m_2 & 0 \cr 0 & -m_1+m_2}
\end{equation}
in a full correspondence with (\ref{77}). Scalar fields $X^{1A}$ and
$X^{2A}$ acquire masses $2m_1$ and $2m_2$ correspondingly, the
spinors $\Omega_i^A$ --- the same masses as the gravitini, while the
other scalar and fermionic fields of $X$-, $Y$- and $Z$-multiplets
remain massless. Also, as one should have expected, all fields of the
vector multiplet except the vector ones acquire masses.

  Moreover, it is interesting that for such a vacuum expectation
values of the scalar fields there exist non-trival Yukawa couplings,
mixing fields from different hypermultiplets:
\begin{equation}
 - \frac12 \bar\Sigma^{i\ddot A} \Gamma^{A\dot A\ddot A}
\varepsilon_{ij} (X^{mA} \bar{\cal Z}^M N^M_m) \Lambda^{j\dot A}.
\end{equation}
Note, that we work in a system where gravitational coupling constant
$k = 1$. The value of Yukawa couplings above, which is determined by
the vacuum expectation value of $\bar{\cal Z}^M N^M_m$, is $m_{1,2} /
m_{pl}$, so for such a coupling to be essential one has to have
the scale of $N=2 \rightarrow N=1$ supersymmetry breaking not much
below the Plank scale, e.g. of the order of the grand unification
scale.

\section*{Conclusion}

  Thus, in this paper we have considered the possibility to switching
on the gauge interaction for both types of $N=2$ supergravity models,
where the scalar fields of the hypermultiplets parameterize the
nonsymmetric quaternionic manifolds. First of all, we were
interested in the possibility to have the spontaneous supersymmetry
breaking without a cosmological term in these models. As we have
shown, for the $W(p,q)$ models the pattern of supersymmetry breaking
resembles very much the one for the usual $O(4,p)/O(4)\otimes O(p)$
quaternionic model. In turn, the spontaneous supersymmetry breaking
in the $V(p,q)$ models leads to a more interesting picture. First,
the mass terms are generated for some of the fields from the
hypermultiplets and not for the vector multiplet ones only. Second,
we have shown that as one of the byproducts of supersymmetry
breaking one obtains the Yukawa couplings for the scalar and spinor
fields of the hypermultiplets. Such couplings, which are absent in
the globally supersymmetric $N=2$ gauge theories as well as in
the $N=2$ supergravity models with the symmetric quaternionic
manifolds, could lead to interesting phenomenological consequences.
We have not considered here a possibility to introduce the nonzero
vacuum expectation values for the matter scalar fields along the flat
directions of the potentials, which could give the gauge symmetry
breaking, because we concentrated here on the general properties of
these models and have not considered any specific models. But the
results already obtained make these models quite promising and
deservs further study.

\vspace{0.3in}
{\large \bf Acknowledgements}
\vspace{0.2in}

The work supported by the International Science Foundation and
Russian Government grant RMP300 and by the Russian Foundation for
Fundamental Research grant 95-02-06312.



\begin{thebibliography}{1}

\bibitem{Tso96}
V.~A. Tsokur and Yu.~M. Zinoviev, $N=2$ supergravity models, based on
the nonsymmetric quaternionic manifolds. I. Symmetries and
Lagrangians, preprint IHEP~96-22, Protvino, 1996, hep-th/9605134.

\bibitem{Ale75}
D.~V. Alekseevskii, {\it Math. USSR Izvestija} {\bf 9} (1975) 297.

\bibitem{Cec89}
S. Cecotti, {\it Com. Math. Phys.} {\bf 124} (1989) 23.

\bibitem{Zin90}
Yu.~M. Zinoviev, {\it Int. J. of Mod. Phys.} {\bf A7} (1992) 7515.

\bibitem{Cec86a}
S. Cecotti, L. Girardello and M. Porrati, {\it Phys. Lett.} {\bf
168B} (1986)
  83.

\bibitem{Zin86}
Yu.~M. Zinoviev, {\it Yad. Fiz.} {\bf 46} (1987) 943.

\bibitem{Wit93}
B. de~Wit, F. Vanderseypen and A. van~Proeyen, {\it Nucl. Phys.} {\bf
B400}   (1993) 463.

\bibitem{Wit94}
B. de~Wit and A. van~Proeyen, {\it Int. J. Mod. Phys.} {\bf D3}
(1994) 31.

\end{thebibliography}
\end{document}